\documentstyle[epsf]{article}
\begin{document}
\bibliographystyle{unsrt}
\title{ Detection of Gravitational Waves from Eccentric Compact
Binaries}
\author{Karl~Martel\\
\small{Department of physics, University of Guelph, Guelph,Ont., N1G
3E1}}
\date{}
\maketitle
\vspace{-0.8cm}
\begin{abstract}
Coalescing compact binaries have been pointed out as the most
promising
source of gravitational waves for kilometer-size interferometers such
as LIGO.  Gravitational wave signals are extracted from the noise in the
detectors by matched filtering.  This technique performs really well if 
an {\it a priori} theoretical knowledge of the signal is
available.  The information known about the possible sources is used to
construct a model of the expected waveforms (templates).  A common
assumption made when constructing templates for coalescing compact
binaries is that the companions move in a quasi-circular orbit.  Some
scenarios, however,  predict the existence of eccentric binaries.  We
investigate the loss in signal-to-noise ratio induced by
non-optimal filtering of eccentric signals.
\end{abstract}

\newcommand{\bvec}[1]{\mbox{\boldmath $#1$\unboldmath}\hspace{-0.3mm}}
\newcommand{\ud}{{\mathrm d}}
\newcommand{\Ibar}{\mathrm{I}\hspace{-0.14cm}\rule[4pt]{1.15mm}{0.2mm}}
\newcommand{\hxx}{\mathrm{h_{+}}}
\newcommand{\hxy}{\mathrm{h_{\times}}} 

\section{Introduction}
Coalescing compact binaries have been pointed out as the most promising
source of gravitational waves for the LIGO/VIRGO/TAMA/GEO
interferometers\cite{kip:300,Schutz:rev}. These binaries
typically have formed a long time ago, giving them time to radiate
most of their eccentricity away.  The templates (model of the radiation)
needed for matched filtering are thus constructed according to this
assumption.  Gravitational waves emitted by a circular binary will be
explicitly searched for in the output of the detectors, but not
gravitational waves emitted by eccentric binaries. The scenario we
have in mind allows
for the formation of young eccentric binary systems, young
enough that they did not have had time to be fully circularized by the
radiation reaction. For example, the collapse of a dense Newtonian
globular cluster can lead to the
formation of a copious number of eccentric binaries via two- and three-
body encounters\cite{ST,SQ}.

These eccentric binaries will emit strongly in the frequency band of the
LIGO interferometers.  It
may seem that these eccentric binaries can be dealt with by
incorporating adequate templates in the bank of templates already 
available, but this may prove to be inefficient.  The addition of new
templates has two undesirable effects: It adds to the already 
heavy computational burden associated with data processing 
and it increases the probability of false detection (mistaking the
noise in the detector for a signal).  A better solution might be to search
for these eccentric signals with the circular templates, and once a signal
is concluded to be present, to extract the information
using eccentric templates.

For this to be possible, the circular templates have to follow
the phase of the eccentric signals very well. To assess the quality of the
circular templates at modeling eccentric signals, special detection tools
are needed.

\section{Matched filtering as a detection method}
Gravitational wave signals are very weak and at best they will be of the
same order of magnitude as the noise in the detectors.  This motivates
the general belief that matched filtering  will be needed to extract the
signals from the noisy output of the detectors \cite{kip:300}.  When the
signals are of known shape, this technique produces the highest
signal-to-noise ratio\cite{HuFla}.  Suppose a gravitational wave $h(t)$
reaches the detector.  The output of the detector $o(t)$ is then a
superposition of the useful signal $h(t)$ and the noise $n(t)$.  In
matched filtering, the signal is extracted by using a theoretical
template (theoretical model) that mimics the signal as well as possible;
we call this template $m(t,\bvec{\Omega})$.  The vector \bvec{\Omega}
denotes the
parameters that characterize the template.  If the template were a perfect
copy of the signal, the parameters would represent the real parameters of 
the source, such as its mass and distance from earth.  If the
templates are not a perfect  approximation to the real signal, the
parameters \bvec{\Omega} represent phenomenological parameters.

If instead of working in the time domain we work in frequency space, we
can introduce the natural inner product of matched filtering.
For two functions $a(t)$ and $b(t)$ with Fourier transforms
$\tilde{a}(f)$ and $\tilde{b}(f)$, the inner product is defined
as\cite{Apostolatos}
\begin{eqnarray}
(a|b)&=&2 \int_{0}^{\infty}\ud f
\frac{\tilde{a}^{*}(f)\tilde{b}(f)+\tilde{a}(f)\tilde{b}^{*}(f)}{S_{n}(f)}\, ,
\end{eqnarray}
where a ``\,\,*\,\,'' denotes complex conjugation and $S_{n}(f)$ is the
one-sided spectral density of the detector's noise.  In terms of this
inner product, the average signal to noise ratio is\cite{wz} 
\begin{eqnarray}
\left<\rho\right>&=&\frac{(m(\bvec{\Omega})|h)}
{\sqrt{(m(\bvec{\Omega})|m(\bvec{\Omega}))}}\, .
\label{eqn:int_snr}
\end{eqnarray}

In practice, the set of parameters \bvec{\Omega} is varied until a maximum
of the signal-to-noise ratio is found.  This maximum is the 
signal-to-noise ratio achievable by using 
$\tilde{m}(f,\bvec{\Omega})$ as a template.  The signal-to-noise ratio of
equation (\ref{eqn:int_snr}) does not give any information about the
quality of the templates or, equivalently, how well the template models
the signal.  The
Schwartz inequality provides an answer to this question\cite{wz}.  The
absolute maximum the signal-to noise ratio can take is achieved when
the template is a {\em perfect} match of the signal, and the
parameters \bvec{\Omega} correspond to the parameters of the source
($\tilde{m}(t,\bvec{\Omega})\equiv \tilde{h}(f)$). The optimal SNR
is\cite{wz}
\begin{eqnarray}
\left<\rho\right>_{max}&=&\sqrt{(h|h)}\, .\label{eqn:snr_max}
\end{eqnarray}
By dividing the signal-to-noise ratio (equation (\ref{eqn:int_snr})) with
the value achieved by optimal filtering (equation (\ref{eqn:snr_max})),
we construct the ambiguity function ${\mathcal A}(\bvec{\Omega})$:
\begin{eqnarray}
{\mathcal A}(\bvec{\Omega}) &=&\frac{(m(\bvec{\Omega})|h)}
{\sqrt{(m(\bvec{\Omega})|m(\bvec{\Omega}))(h|h)}}
\, . \label{eqn:A}
\end{eqnarray}
This function takes values between 0 and 1.  It is equal to 1
when the optimal template is used.

The value of the parameters \bvec{\Omega} can be varied until
${\mathcal A}(\bvec{\Omega})$ is maximized.  The maximum value of the
ambiguity function is the fitting factor:
\begin{eqnarray}
FF=\max_{\bvec{\Omega}}{\mathcal A}(\bvec{\Omega})\, . \label{eqn:FF}
\end{eqnarray}
The fitting factor is a direct measure of
the template's quality since it can be related to the loss of event rate,
i.e. the number of events missed by using an inappropriate set of 
templates. This
loss is calculated according to $1-FF^{3}$\cite{Apostolatos}.  For
example, if the fitting factor is
0.8, then  $48.8\%$ of the events would be mistaken for noise.  We adopt a
threshold of $FF=0.9$ for the present work.  This corresponds to a
loss in event rate of $27\%$.

\section{The gravitational waveforms}
We calculate the waveforms for both circular and eccentric binaries in
the quadrupole approximation. In this approximation the waveforms are
given by\cite{mtw}
\begin{eqnarray}
h^{TT}_{i j}=\frac{2}{R}\frac{\ud^{2}}{\ud t^{2}}\left(I_{i
j}-\frac{1}{3}\delta_{i j}I^{k}\,_{k}\right)^{TT}\, ,
\label{eqn:quadrupole}
\end{eqnarray}  
where $R$ is the distance between the source and the observer, $I_{i
j}$ is the source's quadrupole moment
and the superscript $TT$ reminds us that gravitational waves are
traceless and live in the plane transverse to the direction of
propagation.   

For eccentric binary systems, the waveforms are\cite{Wahlquist}
\begin{eqnarray}
\hxx&=&h_{xx}=-h_{yy}=\frac{1}{R}
\frac{\mu}{p}\Bigg\{2
 (1+\cos^{2}\theta_{o})\cos 2(\varphi-\varphi_{o}) \nonumber \\
 &+& e \left[ (1+\cos^{2}
 \theta_{o})\left(\frac{5}{2}\cos(\varphi-2\varphi_{o})+
 \frac{1}{2}\cos(3\varphi-2\varphi_{o})\right) +\sin^{2}\theta_{o}
 \cos\varphi \right] \nonumber\\
 &+&e^{2}{(1+\cos^{2} \theta_{o})\cos
2\varphi_{o}+\sin^{2}
 \theta_{o}}\Bigg\}\, ,  \label{eqn:h+}\\
\hxy&=&h_{xy}=h_{yx}=-\frac{1}{R}\frac{\mu}{p}
 \cos \theta_{o}\Bigg\{ 4\sin 2(\varphi-2\varphi_{o}) \nonumber \\
 &+&e\left[5\sin (\varphi-2\varphi_{o})+\sin
 (3\varphi-2\varphi_{o})\right\} -2e^{2}\sin 2 \varphi_{o}
\Bigg\}\, ,\label{eqn:hx}
\end{eqnarray}
where $\theta_{o}$ and $\varphi_{o}$ are the two angles defining the
location of the observer with respect to the orbital plane, $\mu$ is the 
reduced mass, and $p$ and $e$ are defined in terms of the turning points
$(r_{\pm})$ of the Newtonian orbit as
\begin{eqnarray*}
r_{\pm}&=&\frac{M p}{1 \pm e}\, .
\end{eqnarray*}
\begin{figure}[!t]
\begin{center}
\epsfxsize=4.5in
\epsfysize=3.0in
\epsfbox{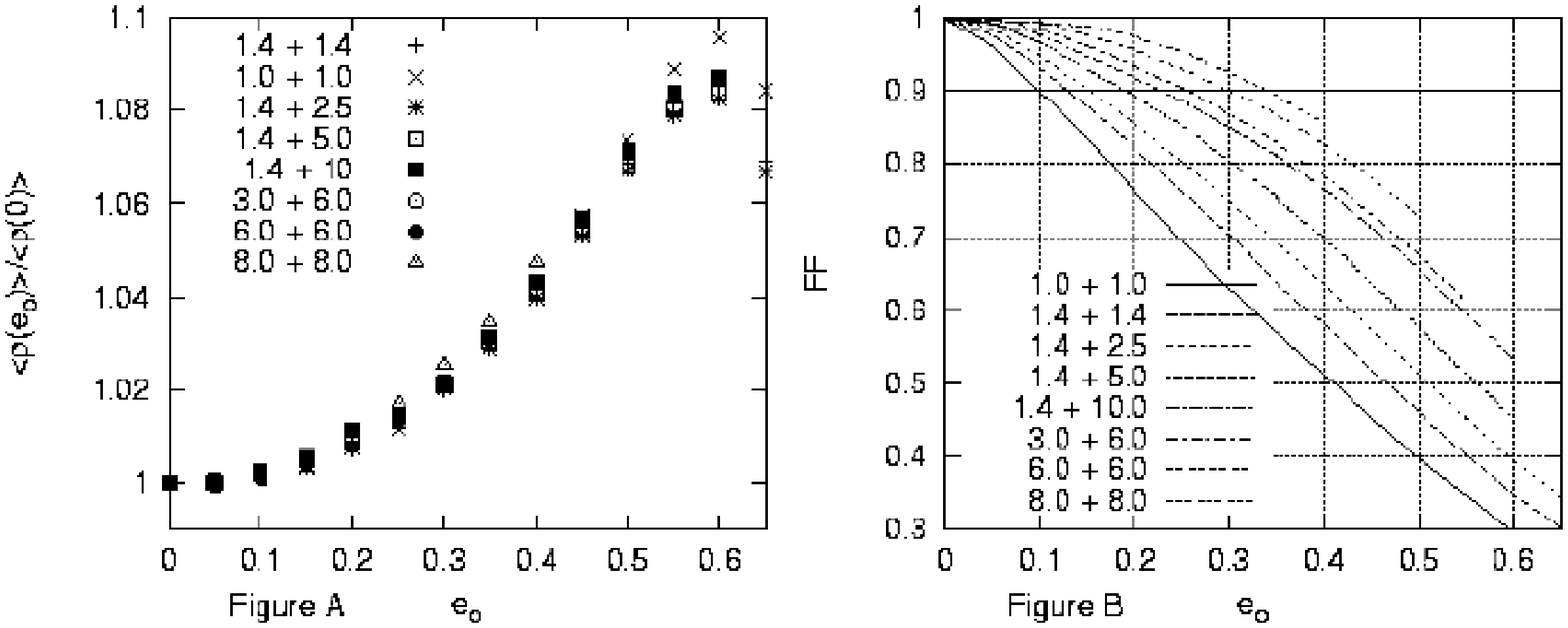}
\caption{Figure A: The ratio of the optimal signal-to-noise ratio
$\left<\rho (e_{o})\right>$ to the signal-to-noise ratio ($\left<\rho
(0)\right>$). 
The figure shows that eccentric binary systems will be easier to
detect if they are explicitly searched for at the output of the
interferometers.\newline\protect 
Figure B: The fitting factor as a function of $e_{o}$
for various binary systems.  Two trends are apparent. The first one is
the net decrease in the fitting factor as $e_{o}$ increases, while the
total mass of the binary is held fixed. The second one is the increase
in the detection probability when the total mass of the
binary increases. The various binaries studied are labeled by the two
masses of the companions; they are given
in units of the solar mass.}\label{fig:fit}
\end{center}
\end{figure}

The eccentric waveforms oscillate at once, twice and thrice the orbital
frequency, whereas the circular waveforms oscillate only at twice the
orbital frequency.  We parameterize our binaries by
specifying the two masses, and the eccentricity $e_{o}$ they have when
they first enter the LIGO frequency band (at 40 Hz).

The Fourier transform of the waveforms is calculated numerically
for the eccentric signal, and obtained through the stationary phase
approximation for the circular templates\cite{FC}.  Once these Fourier
transforms are known, it is
straightforward to calculate the optimal SNR (equation
(\ref{eqn:snr_max})), build the ambiguity function (equation
(\ref{eqn:A})), and  maximize it over the different parameters of the
templates to get the fitting factor (equation (\ref{eqn:FF})). 

The results for the signal-to noise ratio and the fitting factor are
displayed in figure (\ref{fig:fit}).  The signal-to-noise ratio for an
eccentric signal is
higher than the ratio obtained for an equivalent circular binary.
This means that if both binaries are located at the same distance $R$, the
eccentric binary will emit stronger radiation and will be easier to
detect if optimal filters are used. On the other hand, the fitting factor
decreases
as the initial eccentricity is increased. The circular templates fail to
model the eccentric signal properly.  The good news is that circular
templates are still accurate enough to detect some eccentric signals.
For example, a neutron star binary system will be detected as long as its
initial eccentricity does not exceed 0.13.  If the total mass of the
system is increased, the detection probability increases as well.  For
example, for a system of two 8.0\,$M_{\odot}$ black holes, the initial
eccentricity can be as high
as 0.33.  This trend is explained in the following way.  As the total mass
of the system increases, the radiation it emits is stronger and the system
coalesces in a shorter time.  The shorter the signal, the less opportunity
the circular templates have to go out of phase with it.
This is a good news, because the higher the
mass is, the easier it is to detect the signal because it is
stronger.  Thus, the LIGO interferometers
should be able to detect radiation from some eccentric binaries, those
with large masses and relatively low eccentricities\cite{us}.
\newline
\newline
\textbf{Acknowledgments}: This work was carried out with
\'E.~Poisson. It was supported by NSERC.
\bibliography{refs}

\end{document}